\documentclass[aps,pra,reprint,superscriptaddress,twocolumn]{revtex4-2}

\usepackage{graphicx} % Required for inserting images
\usepackage{physics}
\usepackage{braket}
\usepackage{amsmath}
\usepackage{amssymb}
\begin{document}
\title{Frequency resolved optical gating using parametric amplification for characterizing ultrafast temporally multimode squeezed states}

% \author{Elina Sendonaris\authormark{1}, Thomas Zacharias\authormark{2}, Robert Gray \authormark{2}, James Williams\authormark{2}, and Alireza Marandi\authormark{1,2,*}}

% \address{\authormark{1}Department of Applied Physics, California Institute of Technology, Pasadena, 91125, California, USA\\
% \authormark{2}Department of Electrical Engineering, California Institute of Technology, Pasadena, California 91125, USA}

% \email{\authormark{*}marandi@caltech.edu}

\author{Elina Sendonaris}
\affiliation{Department of Applied Physics, California Institute of Technology, Pasadena, 91125, California, USA}

\author{Thomas Zacharias}
\affiliation{Department of Electrical Engineering, California Institute of Technology, Pasadena, California 91125, USA}

\author{Robert Gray}
\affiliation{Department of Electrical Engineering, California Institute of Technology, Pasadena, California 91125, USA}

\author{James Williams}
\affiliation{Department of Electrical Engineering, California Institute of Technology, Pasadena, California 91125, USA}

\author{Alireza Marandi}
\email{marandi@caltech.edu}
\affiliation{Department of Applied Physics, California Institute of Technology, Pasadena, 91125, California, USA}
\affiliation{Department of Electrical Engineering, California Institute of Technology, Pasadena, California 91125, USA}

\begin{abstract}
    Temporally multimode squeezed states have been a topic of recent interest due to their applications in quantum communication, information processing, and sensing. Characterizing the mode shapes is crucial for effectively manipulating these states, but current mode shape and state characterization techniques necessitate constraining assumptions and complicated experimental setups. Here, we propose a characterization technique that simultaneously recovers the complex temporal mode shapes and quadrature variances of ultrafast multimode squeezed states based on frequency resolved optical gating (FROG) using an optical parametric amplifier (OPA). FROG is a promising tool for quantum state characterization due to its flexibility of implementation and high temporal resolution. Using an OPA as the nonlinear process in FROG has the benefit of amplifying weak quantum states to a detectable level while preserving quantum information. Numerical simulations demonstrate the recovery of the mode shapes and levels of squeezing and anti-squeezing of ultrafast multimode squeezed states. This scheme offers a practical experimental approach to measuring arbitrary temporal mode shapes and characterizing large-scale multimode ultrafast  Gaussian quantum states.
\end{abstract}
\maketitle

\section{Introduction}
Optical quantum information processing and communication have been the subjects of intense study, due to light's long coherence time and information density capacity \cite{obrien_optical_2007, asavanant_generation_2019, larsen_deterministic_2019, zhong_quantum_2020, pirandola_advances_2020, larsen_integrated_2025}.
Encoding quantum information in the temporal modes of ultrashort pulses, which are overlapping orthonormal modes in time and frequency, allows for multiple modes to exist within the same pulse and spatial mode \cite{brecht_photon_2015}. This basis allows a single pulse to carry a large amount of information, making it an attractive and natural option for ultrafast time-multiplexed communication and computation \cite{serino_orchestrating_2024}. High-bandwidth single-mode pulses also carry a large amount of information, and recent progress has seen quantum optics extended even to the attosecond regime \cite{cruz-rodriguez_quantum_2024, tzur_attosecond-resolved_2025, tzur2025measuring}.

One difficulty of working with ultrafast quantum states is that the efficiency of operations degrades rapidly with the mismatch between modes, so characterizing the mode shapes of the quantum state is paramount to manipulating these states and realizing useful computation and communication. Using squeezed states to perform continuous-variable quantum computation \cite{menicucci_universal_2006, menicucci_fault-tolerant_2014} and secure quantum communication \cite{ralph_continuous_1999, leverrier_unconditional_2009} is one path to quantum advantage which has the benefit of deterministic state generation. Characterizing ultrafast quantum states has therefore been an active field of study.

In particular, the tomography of temporally multimode squeezed states using optical parametric amplifiers (OPAs) has been the subject of intense recent study \cite{williams_ultrafast_2025-1, barakat_simultaneous_2025, kalash_efficient_2025}. OPAs are well-suited to generating and measuring these states due to their inherently multimode \cite{wasilewski_pulsed_2006} and broadband nature \cite{shaked_lifting_2018}, and their phase-sensitive amplification which allows for quadrature-selective measurement  while preserving some quantum information \cite{nehra_few-cycle_2022, kalash_wigner_2023}. Because of their broad bandwidth \cite{ledezma_intense_2022}, many temporal modes can exist in an ultrashort multimode squeezed pulse generated by an OPA. Mode-selective squeezing measurements allow us to access the squeezing in the principal modes, which is higher than the mode-averaged (e.g. time-averaged) squeezing which is typically measured \cite{ng_quantum_2023, williams_ultrafast_2025-1, barakat_simultaneous_2025}.

One challenge in temporal-mode continuous-variable (CV) quantum optics is that the relevant basis is typically unknown, as the principal squeezing modes depend on pump shape and phase matching and are sensitive to physical imperfections. Most detection strategies require mode-matched local oscillators \cite{cai_multimode_2017} or reconfigurable mode projections \cite{polycarpou_adaptive_2012, karnieli_variational_2025}. Some techniques have a tradeoff between the amount of information gained about the state and the modes \cite{tiedau_quantum_2018}, and others rely on prior knowledge about the state \cite{takase_complete_2019, gil-lopez_universal_2021}. Quantum pulse gates can decompose the state into multiple modes by shaping a pump pulse \cite{serino_realization_2023, serino_orchestrating_2024-1} and have been used to decompose a state into its component modes adaptively \cite{gil-lopez_universal_2021, serino_self-guided_2025}. Electro-optic sampling has also been used to characterize the temporal nature of squeezed light \cite{riek_subcycle_2017, yang_subcycle_2023, hubenschmid_optical_2024, hubenschmid_time-domain_2025}, but it necessitates sub-cycle pulses, which limits it to quantum pulses with longer wavelengths. Lastly, decomposition of data measured using homodyne detection \cite{roslund_wavelength-multiplexed_2014, cai_multimode_2017} can also be used to identify the principal modes of a state. Many of these techniques require pulse shaping and feedback loops, and their resolution and bandwidth is typically limited by pulse shaping performance. However, a technique capable of directly reconstructing both the temporal mode structure and quantum statistics of ultrafast multimode squeezed states without requiring mode-matched local oscillators or adaptive pulse shaping remains a significant challenge.

\begin{figure}
    \centering
    \includegraphics[width=\linewidth]{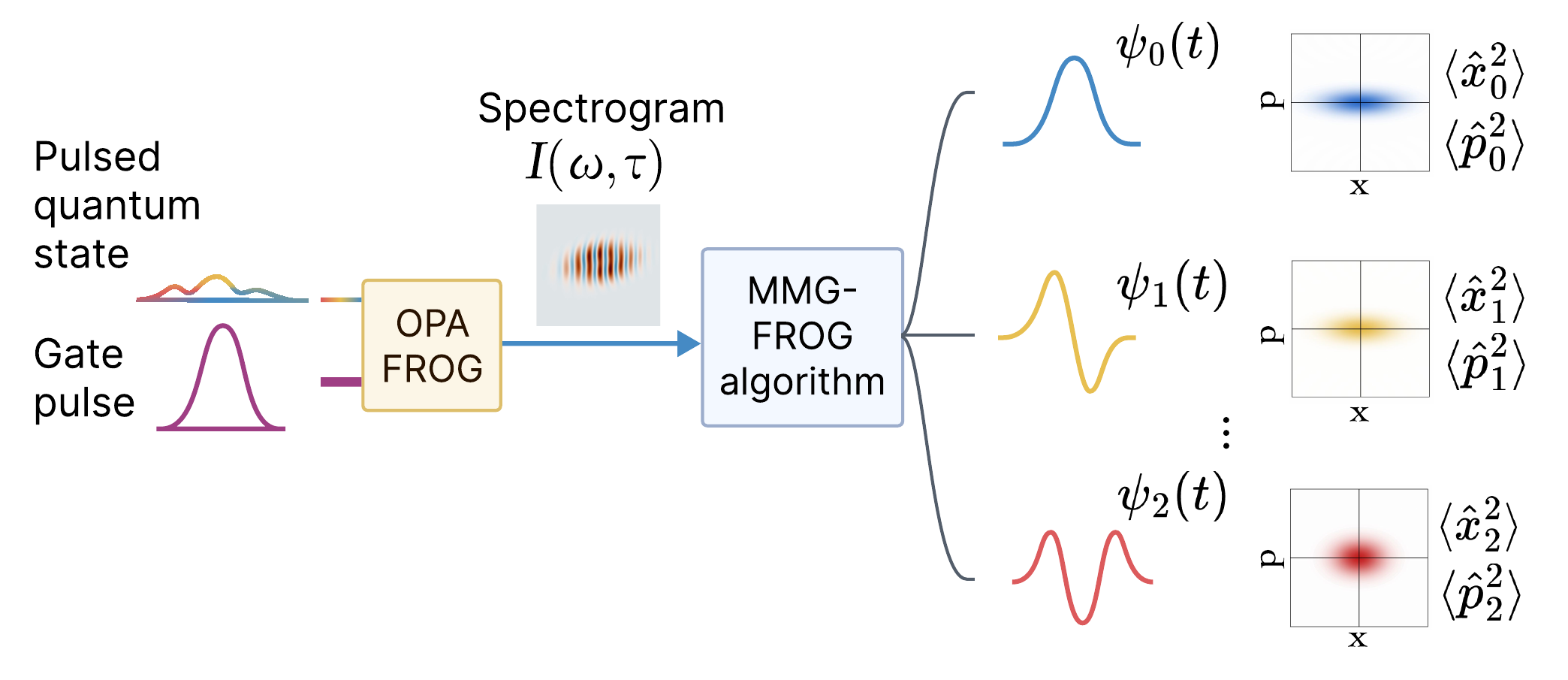}
    \caption{Temporally multimode Gaussian (MMG) state reconstruction process using OPA FROG. The pulsed quantum input state and a gate pulse are put through an OPA FROG setup, and the resulting spectrogram is put into the MMG-OPA-FROG algorithm, which splits the input state up into its component modes $\psi_n(t)$ and second-order moments $\langle \hat{x}_n^2\rangle$ and $\langle \hat{p}_n^2 \rangle$.}
    \label{fig:intro}
\end{figure}

Here we introduce multimode Gaussian OPA frequency-resolved optical gating (MMG-OPA-FROG), a phase-sensitive parametric variant of FROG in which a broadband OPA serves as the nonlinear gate, and show how it can be used to characterize Gaussian states through their quadrature variances and simultaneously determine their principal temporal mode basis. We propose and numerically simulate using an OPA as the nonlinear element in a frequency-resolved optical gating (FROG) setup \cite{trebino_frequency-resolved_2000} and develop a retrieval algorithm to account for the multimode and quantum nature of the input state. The concept is illustrated in Fig. \ref{fig:intro}. 

FROG is popular due to its simplicity of setup and robustness to experimental imperfections while having high resolution and bandwidth. Employing an OPA as the nonlinear process in FROG has been used to measure pulses with extremely low energy \cite{zhang_measurement_2003, zhang_sub_2004, zacharias_energy-efficient_2025}, which is possible because OPAs can amplify pulses by a factor of as large as $10^8$ while having low dispersion. Furthermore, the complex temporal envelope of quantum correlations between twin beams has been measured using FROG \cite{eto_quantum_2024}. Thus, an OPA-based FROG is an ideal candidate for characterizing low-photon-number ultrashort quantum states. The phase-sensitive nature of the OPA provides access to the second-order moments, from which it is possible to fully characterize Gaussian states and even partially characterize non-Gaussian states. This technique gives the full temporal mode information, including the complex mode phase, which is crucial for manipulating the modes. MMG-OPA-FROG can be applied even for single-mode ultrafast quantum states, in order to take advantage of the high resolution and flexibility in wavelength. We have recently explored multimode squeezed state reconstruction using sum-frequency generation FROG \cite{zacharias2025temporal}. The present work instead considers the distinct measurement regime of OPA-only FROG and develops the corresponding formalism for recovering the temporal modes and second-order quadrature moments.

\begin{figure}
    \centering
    \includegraphics[width=0.9\linewidth]{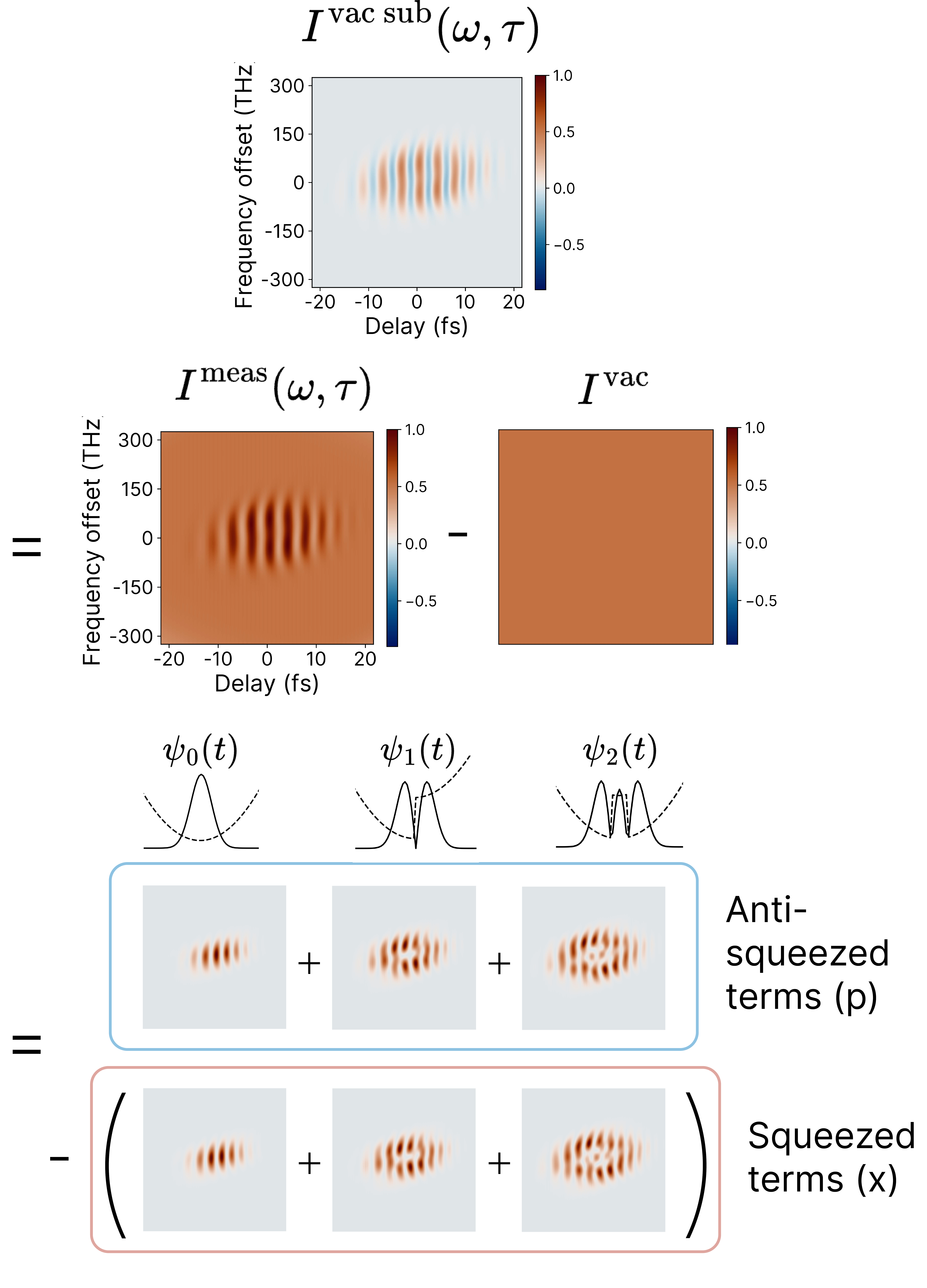}
    \caption{Overview of the vacuum-subtracted spectrogram $I^\text{vac sub}(\omega, \tau)$, which is the measured spectrogram $I^\text{meas}(\omega, \tau)$ minus the vacuum spectrogram $I^\text{vac}$, which is the output when the input state is vacuum. All colorbars are shared and are normalized to the maximum of the measured spectrogram. For multimode squeezed states, the vacuum-subtracted spectrogram is composed of positive elements for the anti-squeezed quadratures and negative elements corresponding to the squeezed quadratures, with two terms per mode (one positive and one negative). Negative values indicate the presence of squeezing.}
    \label{fig:i_vac_sub}
\end{figure}

We show how the FROG spectrogram is related to the mode shapes and second-order moments and then show numerically how ultrafast multimode Gaussian states can be characterized using a FROG retrieval algorithm. In addition, we quantify the resiliency to noise in the spectrogram, showing a successful retrieval even in the presence of noise, and how to extract the relative squeezing angles from the mode shapes. We show how to extract information necessary for temporal mode quantum information processing and communication using a minimal experimental setup and few assumptions.

\section{Theory}
The MMG-OPA-FROG spectrogram is generated by sending the MMG state and gating pulse as the pump into an OPA and measuring the output spectra at various relative delays between the input state and gate. The information in the spectrogram allows for the reconstruction of the complex mode shape of a classical pulse, and it also allows for reconstruction of multiple modes in the same pulse. This spectrogram $I(\omega, \tau)$ is the expectation value of the photon number of the output state at delay $\tau$ and frequency $\omega$
\begin{align}
    I(\omega, \tau) &= \langle (\hat{a}^{\text{out}}(\omega, \tau))^\dag \hat{a}^\text{out} (\omega, \tau)\rangle
\end{align}
where $\hat{a}^\text{out}(\omega, \tau)$ is the annihilation operator just before the spectrometer and the expectation value is over many pulses.

% In the Heisenberg picture, a single-mode OPA transforms the annihilation operator $\hat{a}^\text{in}$ at the input of the OPA to the operator $\hat{a}^\text{out}$ by the Bogoliubov transform \cite{walls_quantum_2008}
% \begin{equation}
% 	\hat{a}^{\text{out}} = \cosh \xi\hat{a}^\text{in} + \sinh \xi (\hat{a}^{\text{in}})^\dag
% \end{equation}
% at the output, for a complex squeezing parameter $\xi$.
% In the case of a multimode OPA, we can define orthonormal temporal modes $\{\psi_m^\text{OPA}(t)\}$ and $\{\phi_m^\text{OPA}(t)\}$ such that for broadband annihilation operators
% \begin{equation}
% 	\hat{a}^\text{out}_m = \int \dd t \psi_m^\text{OPA}(t) \hat{a}^\text{out}(t)
% \end{equation}
% and
% \begin{equation}
% 	\hat{a}^\text{in}_m = \int \dd t \phi_m^\text{OPA}(t) \hat{a}^\text{in}(t),
% \end{equation}
% the action of the OPA can be decomposed into \cite{wasilewski_pulsed_2006}
% \begin{equation}
% 	\hat{a}^\text{out}(t) = \sum_n \psi_m^{\text{OPA}*}(t)(\cosh \xi_n \hat{a}^\text{in}_m + \sinh \xi_n (\hat{a}^\text{in}_m)^\dag),
% \end{equation}
% where $\hat{a}^\text{in/out}(t)$ are the annihilation operators in the time-bin basis.
% Broadband quadrature operators can also be defined:
% \begin{align}
% \hat{x}_m &= \frac{1}{2}(\hat{a}_m + \hat{a}_m^\dag)\\
% \hat{p}_m &= \frac{1}{2i}(\hat{a}_m - \hat{a}_m^\dag)
% \end{align}

We make the following assumptions on the OPA: it is high-gain and it has a much broader bandwidth than the mode basis of the input state. The large-bandwidth approximation, common in FROG reconstruction, means the OPA's principal modes are extremely narrow in time, so we approximate them as delta functions in time. We also assume that the delay $\tau$ is applied to the gating pulse. The annihilation operators at the output of a broadband OPA can then be written as a function of the complex squeezing parameters $\xi(t') =r(t') \exp(i \theta(t'))$ and input state's annihilation operators in the time bin basis $\hat{a}^\text{in}(t)$ as
\begin{equation}
    \hat{a}^\text{out}(t, \tau) =
    \cosh r(t')\hat{a}^\text{in}(t) - e^{i\theta(t')}\sinh r(t')(\hat{a}^{\text{in}}(t))^\dag,
\end{equation}
where $t'\equiv t-\tau$ is the shifted time coordinate. The complex squeezing parameters are dependent on the gating pulse field $E_g(t')$ through $\xi(t') = \kappa E_g(t')$ for nonlinear coupling strength $\kappa$. 

The high-gain approximation entails that, for the highest-gain output modes of the OPA, $r \gg 1$, so $\cosh r \approx \sinh r \approx \exp(r)/2$. We assume that the spectrometer only detects these high-gain modes. We do not make any assumptions on the gating pulse's complex phase, although it must be known.

We now define how we would like to decompose the multimode Gaussian input state. Any Gaussian state can be decomposed into a principal mode basis in which the states are uncorrelated with each other \cite{fabre_modes_2020}. The input state's annihilation operator can be decomposed into its principal mode basis $\{\psi_n(t)\}$ as:
\begin{equation}
    \hat{a}^\text{in}(t) = \sum_n \psi_n^*(t) \hat{a}^\text{in}_n.
\end{equation}
Broadband quadrature operators can also be defined:
\begin{align}
\hat{x}^\text{in}_n &= \frac{1}{2}(\hat{a}^\text{in}_n + \hat{a}^\text{in\dag}_n)\\
\hat{p}^\text{in}_n &= \frac{1}{2i}(\hat{a}^\text{in}_n - \hat{a}^\text{in\dag}_n)
\end{align}

This mode basis is orthonormal, such that $\int \dd t \psi_m(t) \psi^*_n(t) = \delta_{mn}$. In this basis, the expectation values of products of the quadratures of different modes are zero for zero mean field states, such as squeezed states ($\langle \hat{q}_m \hat{q}'_n \rangle \propto \delta_{mn}$ for $\hat{q}, \hat{q}' \in \{\hat{x}^\text{in}, \hat{p}^\text{in}\}$). Since we are focusing on the characterization of multimode squeezed states, we assume the mean-field is zero. For such states, the quadratures can always be rotated such that the covariance $\langle \hat{x}^\text{in}_n\hat{p}^\text{in}_n + \hat{p}^\text{in}_n \hat{x}^\text{in}_n \rangle$ is zero, and the rotation can be absorbed into the mode as a complex phase, so we set that term to zero (see Supplementary Information for further details).

With these assumptions, the spectrogram is
\begin{widetext}
\begin{equation}
\begin{split}
    I(\omega, \tau) = \sum_n \Big[ \langle (\hat{x}_n^\text{in})&^2\rangle \big| \mathfrak{F}\{ G(t') \Im[\psi_n(t) e^{i(\delta\phi(t') + \omega_s \tau)}]\}\big|^2 + \langle (\hat{p}_n^\text{in})^2 \rangle \big| \mathfrak{F}\{ G(t') \Re[\psi_n(t) e^{i(\delta\phi(t') + \omega_s \tau)}]\}\big|^2  \\
    & + \frac{1}{8} \Big( \big| \mathfrak{F}\{G(t') e^{-i\delta\phi(t')}  \psi_n^*(t) \}\big|^2 + \big| \mathfrak{F}\{ G(t') e^{i\delta\phi(t')} \psi_n(t) \}\big|^2 \Big) \Big],
    \label{eqn:spectrogram}
\end{split}
\end{equation}
\end{widetext}
where $\omega_s$ is the central frequency of the input state's modes, $\psi_n(t)$ are the input principal modes, $G(t') = e^{-i\delta\phi(t') + r(t')}$ is a function of the gain and phase of the gating pulse, and $\delta\phi(t') = \theta(t')/2 - \omega_st'$. $\hat{x}_n^\text{in}$ and $\hat{p}_n^\text{in}$ are the quadrature operators for the input state for mode $n$.

\begin{figure*}
    \centering
    \includegraphics[width=\linewidth]{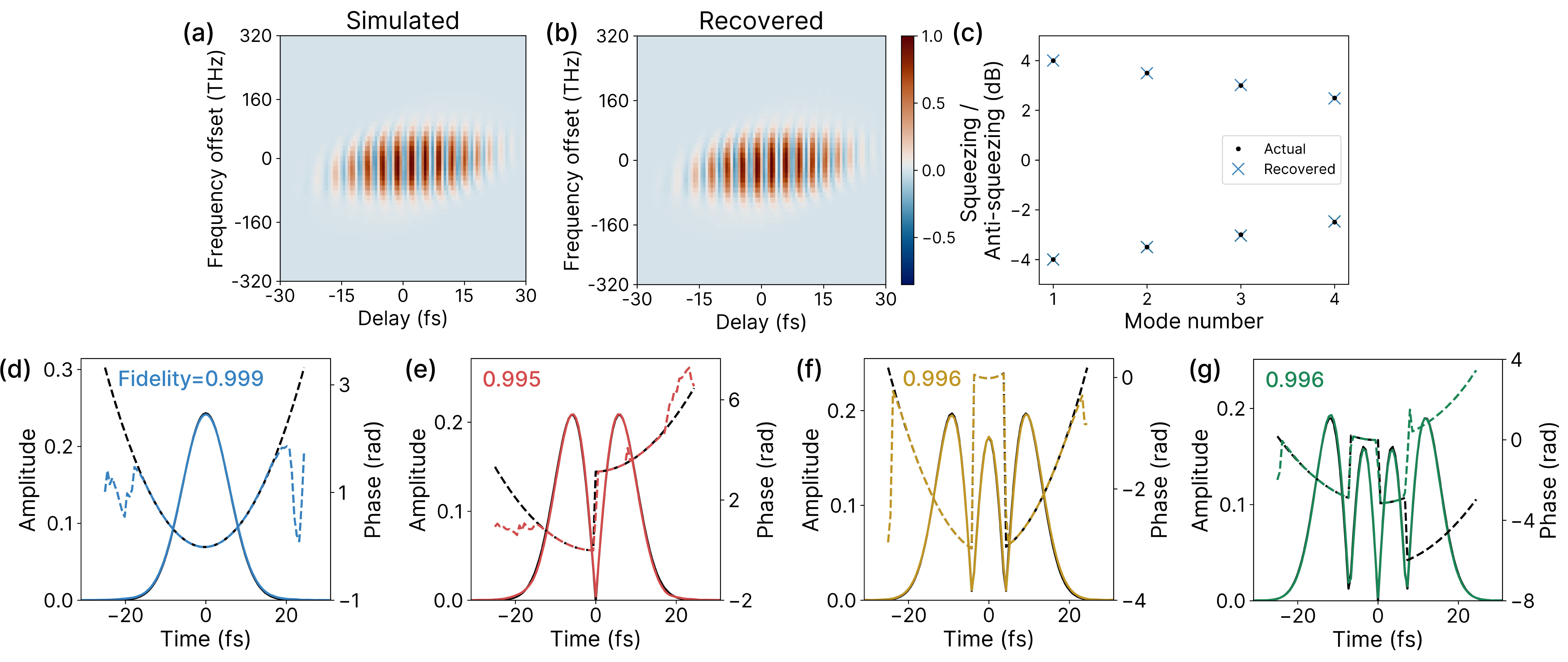}
    \caption{Simulated retrieval of an ultrafast multimode squeezed state using MMG-OPA-FROG algorithm using a 100-fs chirped Gaussian gating pulse. (a) Simulated spectrogram, normalized such that the peak value is 1. (b) Retrieved spectrogram, also normalized. The root mean square distance loss is 0.0026. (c)  Plot of actual (black circles) and recovered (blue crosses) squeezing and anti-squeezing for each mode. (d)-(g) Recovered (colored) and actual (black) mode shapes, with the solid lines portraying amplitude and the dashed lines phase. The fidelity between the actual and recovered modes is written in each plot.}
    \label{fig:results}
\end{figure*}

This form of the spectrogram is specific to bases in which the states in the modes are uncorrelated, so finding the modes which contribute in this way to the spectrogram will give us the principal modes, effectively diagonalizing the covariance matrix. The spectrogram is the incoherent sum of the principal-mode spectrograms, with four terms per mode. Two of the terms are weighted by the two non-zero second-order moments, and the last two are independent of the state's statistics. The angle in quadrature phase space along which the state is squeezed is encoded in the relative phase between the modes, allowing us to recover the covariance of the squeezed state.

For the case of a vacuum input, the spectrogram simplifies to the constant $I^\text{vac}(\omega, \tau) = |\int \dd t~G(t)|^2/4$, which is just the optical parametric generation from the OPA. Because the spectrogram is non-zero for vacuum input, it is possible to extract the absolute squeezing parameters by normalizing the extracted mode shapes using the vacuum spectrogram as a reference. Furthermore, if the vacuum spectrogram is subtracted from a measured spectrogram, the last two terms and all contributions from un-squeezed modes in the input state are canceled out, truncating the sum to only those modes which are occupied by a non-vacuum state. This spectrogram $I^\text{vac sub}(\omega, \tau) \equiv I(\omega, \tau) - I^\text{vac}(\omega, \tau)$, which we call the vacuum-subtracted spectrogram, has negative components if either quadrature has noise below the shot noise level. The negativity of the vacuum-subtracted spectrogram thus gives an indication of whether there is squeezing. An example vacuum-subtracted spectrogram for a multimode squeezed state is shown in Fig. \ref{fig:i_vac_sub} using 25-fs Hermite-Gaussian modes with a moderate chirp and a 100-fs chirped gating pulse. The modes have 3 dB, 4 dB, and 2 dB of squeezing, in order of increasing Hermite polynomial order.

\section{Results}
We numerically simulated the generation of the OPA FROG spectrogram and retrieved the mode shapes and quadrature variances using an MMG-OPA-FROG algorithm. We developed a retrieval algorithm similar to the one in \cite{bourassin-bouchet_partially_2015} to extract the shape of modes, the variances in $\hat{x}$ and $\hat{p}$, and the relative angles of squeezing. We adapted the data constraint to handle negative terms in the vacuum-subtracted spectrogram using Lagrangian multipliers. We used gradient descent to enforce the nonlinear constraint, which ensures that the spectrogram is indeed in the form of Eq.~\ref{eqn:spectrogram}. We use one gradient term for the $\hat{x}^2$ and $\hat{p}^2$ terms for each mode to update the estimates for the mode shapes and variances, followed by a Gram-Schmidt decomposition to ensure mode orthonormality. Further details of the algorithm are in the Supplemental Information.

We numerically generated the spectrogram from a multimode squeezed state composed of 30-fs-long chirped Hermite Gaussian temporal modes, which are a good approximation for the modes that arise from a broadband OPA \cite{wasilewski_pulsed_2006}, using Eq. \ref{eqn:spectrogram}. The gating pulse is a mildly chirped 100-fs Gaussian pulse which amplifies the state by 50 dB. The results, after 10,000 iterations of the algorithm, are shown in Fig. \ref{fig:results}. The loss, which is the root mean square distance between the simulated and recovered spectrograms normalized by the root mean square of the measured spectrogram, is 0.0026.

The mode fidelity of each mode, written as an inset in Figs.~\ref{fig:results}(d)-(g), which we define as the squared overlap between the actual and recovered modes as $|\int \dd t~\psi_n^*(t)\psi^\text{rec}_n(t)|^2$, is above 0.995 for all modes. The amount of actual and recovered squeezing and anti-squeezing is shown in Fig.~\ref{fig:results}(c), with all recovered values falling within 2\% of the actual values. All of these quantities indicate a successful reconstruction of the state. This demonstrates that the MMG-OPA-FROG algorithm can accurately recover both the temporal structure and quantum statistics of ultrafast multimode squeezed states directly from the measured spectrogram.

The algorithm can identify the relative phases, which is equivalent to the squeezing angles for a multimode squeezed state, which can be used to reconstruct the full covariance matrix. In this case the relative phases are zero. Recovery of arbitrary relative phase is shown in the Supplementary Information.

\begin{figure*}
    \centering
    \includegraphics[width=\linewidth]{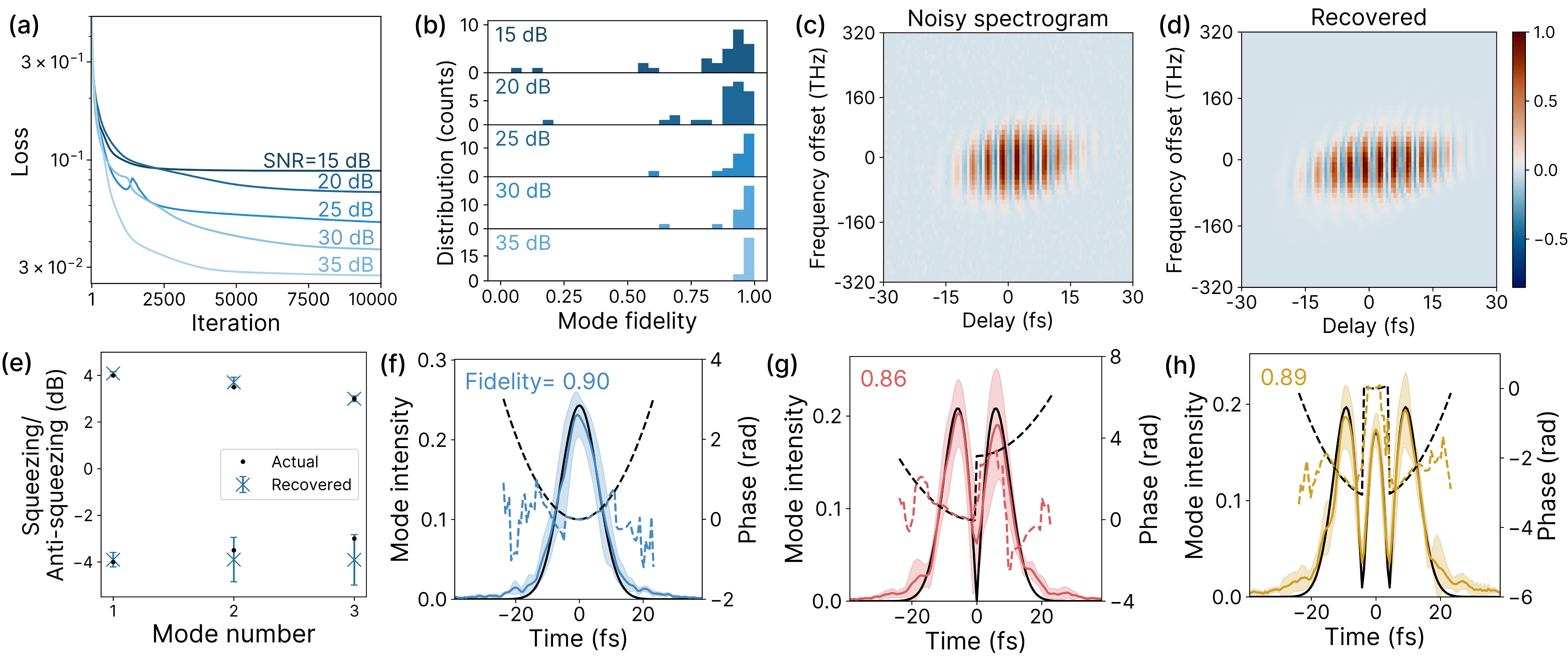}
    \caption{Performance of the algorithm in the presence of noise. (a) Loss vs. iteration number for different amounts of noise in the spectrogram. (b) Distributions of mode fidelity for different SNRs. (c) Example noisy spectrogram with 15 dB SNR. (d) Recovered spectrogram, with the noisy one as input. (e) Actual (dots) and recovered (crosses) squeezing and anti-squeezing values from the noisy spectrogram. (f-h) Recovered mode intensity (solid lines) and phase (dashed lines) from the noisy spectrogram, averaged over successful retrievals. The shaded region indicates one standard deviation, based on bootstrapping. The fidelity of the recovered mode with the actual mode is inset.}
    \label{fig:modes_snr}
\end{figure*}

% This algorithm can also reveal the number of squeezed modes present in a state by sweeping the number of modes in reconstruction until a minimum is reached. This trend can be seen in Fig. \ref{fig:modes_snr}(a), where the loss is high when the number of modes recovered during reconstruction is less than the number of modes in the spectrogram.

We characterize the noise-tolerance of the algorithm by adding Gaussian noise to the simulated spectrogram to achieve various signal-to-noise ratios (SNRs) and characterizing the loss and mode fidelity, as shown in Fig. \ref{fig:modes_snr}(a) and (b). We employ a zeroing mask over the data outside the significantly non-zero portion of the spectrogram so that the noise in the wings does not prevent convergence. In order to simulate a realistic recovery procedure, we use the common bootstrapping technique to generate the error bars from a single noisy spectrogram at each noise level \cite{wang_beyond_2003}. We can see that the recovery decreases in effectiveness with higher noise, and that some retrievals fail (note the heavy tails of the fidelity distributions towards low fidelity for low SNR in Fig. \ref{fig:modes_snr}(b)). These tails indicate that some retrievals fail, so we set a threshold of 10\% loss for considering a state recovery successful, and the following plots are based on successful recoveries.

We show an example mode retrieval at an SNR of 15 dB in Figs. \ref{fig:modes_snr}(c-h). The mode retrieval is slightly less successful than in the noiseless case, with the squeezing values in particular having a large variance. However, the average mode fidelities, shown in the insets of Figs. \ref{fig:modes_snr}(f-h), are above 0.85, and the mode variances are within 10\% of their actual values, showing that the mode shape reconstruction was successful even in the presence of noise.

% \begin{figure}
%     \centering
%     \includegraphics[width=\linewidth]{fig_4.pdf}
%     \caption{(a) Loss vs. number of recovered modes for different numbers of modes in the spectrogram. (b) Loss vs. iteration number for different SNRs, averaged over 10 bootstrap instances.}
%     \label{fig:modes_snr}
% \end{figure}

\section{Discussion}

\begin{figure}[t]
    \centering
    \includegraphics[width=\linewidth]{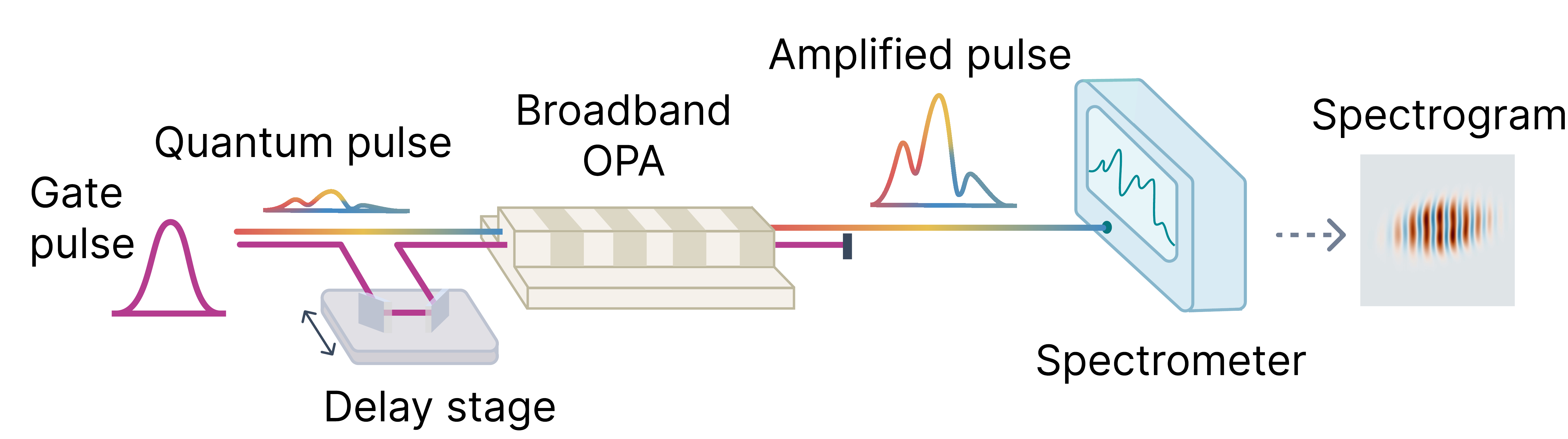}
    \caption{An experimental setup to measure the OPA FROG multimode spectrogram. The quantum pulse is amplified by a delayed gating pulse in a broadband high-gain OPA. The resulting amplified pulse is measured in a spectrometer and these spectra are combined to form the spectrogram $I(\omega, \tau)$.}
    \label{fig:exp}
\end{figure}

The experimental prospects for generating and characterizing a multimode squeezed state using MMG-OPA-FROG are promising. A potential setup to implement it is shown in Fig. \ref{fig:exp}. Multimode squeezed states have been generated and characterized using OPAs \cite{williams_ultrafast_2025-1, barakat_simultaneous_2025, kalash_efficient_2025}, and OPAs have been used as the nonlinear process in FROG setups for classical pulses \cite{zhang_measurement_2003, zhang_sub_2004, zacharias_energy-efficient_2025}. OPAs with a broad bandwidth, which is necessary for the spectrogram to be in this form, are available in integrated photonics \cite{ledezma_intense_2022}. Furthermore, the length of the pump pulse does not limit the resolution of the reconstruction, within reason, and so the pump pulse can be any pulse that allows for broadband and high-gain operation of the OPA, lifting the pulse-shaping requirement present in homodyne detection.

One difficulty of implementing this state retrieval scheme is minimizing the loss between the state-generation and measurement OPA, so that the squeezing is maintained. One solution is putting both OPAs on one integrated photonic chip, which reduces loss and allows for significant squeezing to be measured \cite{nehra_few-cycle_2022}. Another potential difficulty lies in stabilizing the relative delay between the state and gating pulse, since the amplified state coming out of an OPA is very sensitive to the relative phase of the pulses and as such it must be very well stabilized. There are, however, techniques for calibrating the phase given instabilities \cite{zacharias_energy-efficient_2025}.

Currently, except for the mode orthonormalization step, the computation time of the algorithm scales linearly with the number of modes and with the number of data points, since each data constraint and gradient descent step must be implemented mode-by-mode and data-point by data-point, leading to a scaling of $\mathcal{O}(MN_\omega N_\tau)$ for $M$ the number of modes recovered, $N_\omega$ the number of frequency data points, and $N_\tau$ the number of delays measured. The mode orthonormalization, which is a Gram-Schmidt decomposition, scales as $\mathcal{O}(N_\omega M^2)$, but in practice up to $M=40$ modes the linear factor dominates. For retrieving five modes using 10,000 iterations, twelve minutes are required on an Apple M1 processor with no GPU speedup. Due to the linear scaling, by optimizing the algorithm further and using dedicated hardware it may be possible to scale up the retrieval to a larger number of modes with a reasonable computation time. 

Further work includes advancing the measurement scheme, including the covariance. The covariance term is not in the typical form of a FROG spectrogram, so the algorithm must be adjusted to account for this. The covariance would grant us access to some information on non-Gaussian states \cite{yang_subcycle_2023}. To characterize non-Gaussian states fully, we require access to all of the statistics of the state, not just the second-order moments. If a single-shot spectrometer is available, it may be possible to perform full multimode state tomography \cite{gil-lopez_universal_2021}.

\section{Conclusion}
In summary, we have proposed and numerically simulated the characterization of a multimode squeezed state using OPA FROG. We derived the OPA FROG spectrogram of a multimode quantum state as a function of the second-order moments and mode shapes. We developed a multimode FROG algorithm to extract the mode shapes and $\hat{x}$ and $\hat{p}$ variances. This technique can be implemented using available experimental techniques and equipment, particularly the scalable platform of integrated photonics, and has the potential to characterize a wide range of mode shapes and states. Unlike existing techniques, MMG-OPA-FROG does not have demanding gating pump pulse requirements, and there are few limits on the range of wavelengths that can be characterized. There are no assumptions on the shapes of the modes or on the gating pulse, and the assumption that the OPA used in FROG is broadband and high-gain is consistent with current experiments. Tomography and access to the covariance for states beyond squeezed vacuum may become available by further adapting the algorithm.

The ability to characterize ultrafast temporally multimode Gaussian states is central to implementing an optical temporal-mode-based communication and computation system, since knowing the mode shapes is crucial for mode matching and manipulation.  This technique is one possible tool in the toolbox of ultrafast optical temporal mode computing and communication.

\section{Acknowledgments}
E.S. would like to thank Dr. Jamison Sloan for helpful discussions.

\section{Funding}
The authors gratefully acknowledge support from ARO grant no. W911NF-23-1-0048, NSF grants no. 1918549 and 2139433, AFOSR award FA9550-23-1-0755, DARPA award D23AP00158, the center for sensing to intelligence at Caltech, the Alfred P. Sloan Foundation, and NASA/JPL. 

\section{Disclosures}
ES, TZ, RG, AM: Provisional patent application CIT-9310-P (P)
AM: PINC Technologies Inc. (I, E, P)

\section{Data Availability}
The data that support the findings of this study are available from the corresponding author upon reasonable request.

\bibliography{references-2}

@inproceedings{zacharias2025temporal,
  title={Temporal Characterization of Ultrashort-Pulse Squeezed Vacuum using Frequency Resolved Optical Gating},
  author={Zacharias, Thomas and Gray, Robert and Sendonaris, Elina and Williams, James and Shen, Maximilian and Zhou, Selina and Marandi, Alireza},
  booktitle={2025 Conference on Lasers and Electro-Optics (CLEO)},
  pages={1--2},
  year={2025},
  organization={IEEE}
}

@article{tzur2025measuring,
  title={Measuring and controlling the birth of quantum attosecond pulses},
  author={Tzur, Matan Even and Mor, Chen and Yaffe, Noa and Birk, Michael and Rasputnyi, Andrei and Kneller, Omer and Nisim, Ido and Kaminer, Ido and Kr{\"u}ger, Michael and Dudovich, Nirit and others},
  journal={arXiv preprint arXiv:2502.09427},
  year={2025}
}

@article{serino_self-guided_2025,
	year = {2025},
	month = {4},
	title = {Self-guided tomography of time-frequency qudits},
	volume = {10},
	issn = {2058-9565},
	url = {https://iopscience.iop.org/article/10.1088/2058-9565/adb0ea},
	doi = {10.1088/2058-9565/adb0ea},
	pages = {025024},
	number = {2},
	journal = {Quantum Science and Technology},
	shortjournal = {Quantum Sci. Technol.},
	author = {Serino, Laura and Rambach, Markus and Brecht, Benjamin and Romero, Jacquiline and Silberhorn, Christine},
	urldate = {2026-03-03},
	date = {2025-04-01},
	langid = {english},
}

@article{gil-lopez_universal_2021,
	year = {2021},
	month = {10},
	title = {Universal compressive tomography in the time-frequency domain},
	volume = {8},
	rights = {© 2021 Optical Society of America},
	issn = {2334-2536},
	url = {https://opg.optica.org/optica/abstract.cfm?uri=optica-8-10-1296},
	doi = {10.1364/OPTICA.427645},
	pages = {1296--1305},
	number = {10},
	journal = {Optica},
	shortjournal = {Optica, {OPTICA}},
	publisher = {Optica Publishing Group},
	author = {Gil-Lopez, Jano and Teo, Yong Siah and De, Syamsundar and Brecht, Benjamin and Jeong, Hyunseok and Silberhorn, Christine and Sánchez-Soto, Luis L.},
	urldate = {2026-02-18},
	date = {2021-10-20},
}

@article{wang_beyond_2003,
	year = {2003},
	month = {12},
	title = {Beyond error bars: Understanding uncertainty in ultrashort-pulse frequency-resolved-optical-gating measurements in the presence of ambiguity},
	volume = {11},
	rights = {https://doi.org/10.1364/{OA}\_License\_v1\#{VOR}-{OA}},
	issn = {1094-4087},
	url = {https://opg.optica.org/abstract.cfm?URI=oe-11-26-3518},
	doi = {10.1364/OE.11.003518},
	shorttitle = {Beyond error bars},
	pages = {3518},
	number = {26},
	journal = {Optics Express},
	shortjournal = {Opt. Express},
	author = {Wang, Ziyang and Zeek, Erik and Trebino, Rick and Kvam, Paul},
	urldate = {2026-02-12},
	date = {2003-12-29},
	langid = {english},
}

@article{williams_ultrafast_2025-1,
	year = {2025},
	month = {10},
	title = {Ultrafast all-optical measurement of squeezed vacuum in a lithium niobate nanophotonic circuit},
	volume = {7},
	issn = {2643-1564},
	url = {https://link.aps.org/doi/10.1103/rvmb-ljd3},
	doi = {10.1103/rvmb-ljd3},
	pages = {043006},
	number = {4},
	journal = {Physical Review Research},
	shortjournal = {Phys. Rev. Research},
	author = {Williams, James and Sendonaris, Elina and Nehra, Rajveer and Gray, Robert M. and Sekine, Ryoto and Ledezma, Luis and Marandi, Alireza},
	urldate = {2026-01-29},
	date = {2025-10-01},
	langid = {english},
}

@misc{tzur_attosecond-resolved_2025,
	year = {2025},
	month = {11},
	title = {Attosecond-resolved quantum fluctuations of light and matter},
	url = {http://arxiv.org/abs/2511.18362},
	doi = {10.48550/arXiv.2511.18362},
	number = {{arXiv}:2511.18362},
	publisher = {{arXiv}},
	author = {Tzur, Matan Even and Mor, Chen and Yaffe, Noa and Birk, Michael and Rasputnyi, Andrei and Kneller, Omer and Nisim, Ido and Kaminer, Ido and Chekhova, Maria and Krueger, Michael and Ivanov, Misha and Dudovich, Nirit and Cohen, Oren},
	urldate = {2026-01-15},
	date = {2025-11-23},
	eprinttype = {arxiv},
	eprint = {2511.18362 [physics]},
}

@misc{karnieli_variational_2025,
	year = {2025},
	month = {9},
	title = {Variational processing of multimode squeezed light},
	url = {http://arxiv.org/abs/2509.16753},
	doi = {10.48550/arXiv.2509.16753},
	number = {{arXiv}:2509.16753},
	publisher = {{arXiv}},
	author = {Karnieli, Aviv and Mor, Paul-Alexis and Roques-Carmes, Charles and Lustig, Eran and Sloan, Jamison and Vučković, Jelena and Miller, David A. B. and Fan, Shanhui},
	urldate = {2026-01-11},
	date = {2025-09-20},
	eprinttype = {arxiv},
	eprint = {2509.16753 [quant-ph]},
}

@book{trebino_frequency-resolved_2000,
	year = {2000},
	location = {Boston, {MA}},
	title = {Frequency-Resolved Optical Gating: The Measurement of Ultrashort Laser Pulses},
	rights = {http://www.springer.com/tdm},
	isbn = {978-1-4613-5432-1 978-1-4615-1181-6},
	url = {http://link.springer.com/10.1007/978-1-4615-1181-6},
	doi = {10.1007/978-1-4615-1181-6},
	shorttitle = {Frequency-Resolved Optical Gating},
	publisher = {Springer {US}},
	author = {Trebino, Rick},
	urldate = {2025-10-28},
	date = {2000},
	langid = {english},
}

@article{polycarpou_adaptive_2012,
	year = {2012},
	title = {Adaptive Detection of Arbitrarily Shaped Ultrashort Quantum Light States},
	volume = {109},
	doi = {10.1103/PhysRevLett.109.053602},
	number = {5},
	journal = {Physical Review Letters},
	shortjournal = {Phys. Rev. Lett.},
	author = {Polycarpou, C.},
	date = {2012},
}

@article{larsen_integrated_2025,
	year = {2025},
	month = {6},
	title = {Integrated photonic source of Gottesman–Kitaev–Preskill qubits},
	volume = {642},
	rights = {2025 The Author(s)},
	issn = {1476-4687},
	url = {https://www.nature.com/articles/s41586-025-09044-5},
	doi = {10.1038/s41586-025-09044-5},
	pages = {587--591},
	number = {8068},
	journal = {Nature},
	publisher = {Nature Publishing Group},
	author = {Larsen, M. V. and Bourassa, J. E. and Kocsis, S. and Tasker, J. F. and Chadwick, R. S. and González-Arciniegas, C. and Hastrup, J. and Lopetegui-González, C. E. and Miatto, F. M. and Motamedi, A. and Noro, R. and Roeland, G. and Baby, R. and Chen, H. and Contu, P. and Di Luch, I. and Drago, C. and Giesbrecht, M. and Grainge, T. and Krasnokutska, I. and Menotti, M. and Morrison, B. and Puviraj, C. and Rezaei Shad, K. and Hussain, B. and {McMahon}, J. and Ortmann, J. E. and Collins, M. J. and Ma, C. and Phillips, D. S. and Seymour, M. and Tang, Q. Y. and Yang, B. and Vernon, Z. and Alexander, R. N. and Mahler, D. H.},
	urldate = {2025-10-01},
	date = {2025-06},
	langid = {english},
}

@article{zacharias_energy-efficient_2025,
	year = {2025},
	month = {3},
	title = {Energy-Efficient Ultrashort-Pulse Characterization Using Nanophotonic Parametric Amplification},
	volume = {12},
	url = {https://doi.org/10.1021/acsphotonics.4c02620},
	doi = {10.1021/acsphotonics.4c02620},
	pages = {1316--1320},
	number = {3},
	journal = {{ACS} Photonics},
	shortjournal = {{ACS} Photonics},
	publisher = {American Chemical Society},
	author = {Zacharias, Thomas and Gray, Robert and Sekine, Ryoto and Williams, James and Zhou, Selina and Marandi, Alireza},
	urldate = {2025-09-03},
	date = {2025-03-19},
}

@article{menicucci_fault-tolerant_2014,
	year = {2014},
	month = {3},
	title = {Fault-Tolerant Measurement-Based Quantum Computing with Continuous-Variable Cluster States},
	volume = {112},
	url = {https://link.aps.org/doi/10.1103/PhysRevLett.112.120504},
	doi = {10.1103/PhysRevLett.112.120504},
	pages = {120504},
	number = {12},
	journal = {Physical Review Letters},
	shortjournal = {Phys. Rev. Lett.},
	publisher = {American Physical Society},
	author = {Menicucci, Nicolas C.},
	urldate = {2025-08-15},
	date = {2014-03-26},
}

@article{zhang_measurement_2003,
	year = {2003},
	month = {3},
	title = {Measurement of the intensity and phase of attojoule femtosecond light pulses using Optical-Parametric-Amplification Cross-Correlation Frequency-Resolved Optical Gating},
	volume = {11},
	rights = {https://doi.org/10.1364/{OA}\_License\_v1\#{VOR}-{OA}},
	issn = {1094-4087},
	url = {https://opg.optica.org/oe/abstract.cfm?uri=oe-11-6-601},
	doi = {10.1364/OE.11.000601},
	pages = {601},
	number = {6},
	journal = {Optics Express},
	shortjournal = {Opt. Express},
	author = {Zhang, Jing-yuan and Shreenath, Aparna and Kimmel, Mark and Zeek, Erik and Trebino, Rick and Link, Stephan},
	urldate = {2025-08-14},
	date = {2003-03-24},
	langid = {english},
}

@article{barakat_simultaneous_2025,
	year = {2025},
	month = {2},
	title = {Simultaneous measurement of multimode squeezing through multimode phase-sensitive amplification},
	volume = {3},
	rights = {\&\#169; 2025 Optica Publishing Group},
	issn = {2837-6714},
	url = {https://opg.optica.org/opticaq/abstract.cfm?uri=opticaq-3-1-36},
	doi = {10.1364/OPTICAQ.524682},
	pages = {36--44},
	number = {1},
	journal = {Optica Quantum},
	shortjournal = {Optica Quantum, {OPTICAQ}},
	publisher = {Optica Publishing Group},
	author = {Barakat, Ismail and Kalash, Mahmoud and Scharwald, Dennis and Sharapova, Polina and Lindlein, Norbert and Chekhova, Maria},
	urldate = {2025-08-13},
	date = {2025-02-25},
}

@misc{kalash_efficient_2025,
	year = {2025},
	month = {8},
	title = {Efficient detection of spectrally multimode squeezed light through optical parametric amplification},
	url = {http://arxiv.org/abs/2508.04502},
	doi = {10.48550/arXiv.2508.04502},
	number = {{arXiv}:2508.04502},
	publisher = {{arXiv}},
	author = {Kalash, Mahmoud and Han, Ui-Nyung and Ra, Young-Sik and Chekhova, Maria V.},
	urldate = {2025-08-11},
	date = {2025-08-06},
	eprinttype = {arxiv},
	eprint = {2508.04502 [quant-ph]},
}

@article{bourassin-bouchet_partially_2015,
	year = {2015},
	month = {3},
	title = {Partially coherent ultrafast spectrography},
	volume = {6},
	rights = {2015 The Author(s)},
	issn = {2041-1723},
	url = {https://www.nature.com/articles/ncomms7465},
	doi = {10.1038/ncomms7465},
	pages = {6465},
	number = {1},
	journal = {Nature Communications},
	shortjournal = {Nat Commun},
	publisher = {Nature Publishing Group},
	author = {Bourassin-Bouchet, C. and Couprie, M.-E.},
	urldate = {2025-04-23},
	date = {2015-03-06},
	langid = {english},
}

@article{hubenschmid_optical_2024,
	year = {2024},
	month = {11},
	title = {Optical Time-Domain Quantum State Tomography on a Subcycle Scale},
	volume = {14},
	issn = {2160-3308},
	url = {https://link.aps.org/doi/10.1103/PhysRevX.14.041032},
	doi = {10.1103/PhysRevX.14.041032},
	pages = {041032},
	number = {4},
	journal = {Physical Review X},
	shortjournal = {Phys. Rev. X},
	author = {Hubenschmid, Emanuel and Guedes, Thiago L. M. and Burkard, Guido},
	urldate = {2025-03-21},
	date = {2024-11-05},
	langid = {english},
}

@article{tiedau_quantum_2018,
	year = {2018},
	month = {3},
	title = {Quantum state and mode profile tomography by the overlap},
	volume = {20},
	issn = {1367-2630},
	url = {https://dx.doi.org/10.1088/1367-2630/aaad8a},
	doi = {10.1088/1367-2630/aaad8a},
	pages = {033003},
	number = {3},
	journal = {New Journal of Physics},
	shortjournal = {New J. Phys.},
	publisher = {{IOP} Publishing},
	author = {Tiedau, J and Shchesnovich, V S and Mogilevtsev, D and Ansari, V and Harder, G and Bartley, T J and Korolkova, N and Silberhorn, Ch},
	urldate = {2025-03-01},
	date = {2018-03},
	langid = {english},
}

@article{takase_complete_2019,
	year = {2019},
	month = {3},
	title = {Complete temporal mode characterization of non-Gaussian states by a dual homodyne measurement},
	volume = {99},
	url = {https://link.aps.org/doi/10.1103/PhysRevA.99.033832},
	doi = {10.1103/PhysRevA.99.033832},
	pages = {033832},
	number = {3},
	journal = {Physical Review A},
	shortjournal = {Phys. Rev. A},
	publisher = {American Physical Society},
	author = {Takase, Kan and Okada, Masanori and Serikawa, Takahiro and Takeda, Shuntaro and Yoshikawa, Jun-ichi and Furusawa, Akira},
	urldate = {2025-02-28},
	date = {2019-03-18},
}

@article{menicucci_universal_2006,
	year = {2006},
	month = {9},
	title = {Universal Quantum Computation with Continuous-Variable Cluster States},
	volume = {97},
	rights = {http://link.aps.org/licenses/aps-default-license},
	issn = {0031-9007, 1079-7114},
	url = {https://link.aps.org/doi/10.1103/PhysRevLett.97.110501},
	doi = {10.1103/PhysRevLett.97.110501},
	pages = {110501},
	number = {11},
	journal = {Physical Review Letters},
	shortjournal = {Phys. Rev. Lett.},
	author = {Menicucci, Nicolas C. and Van Loock, Peter and Gu, Mile and Weedbrook, Christian and Ralph, Timothy C. and Nielsen, Michael A.},
	urldate = {2025-02-05},
	date = {2006-09-13},
	langid = {english},
}

@article{roslund_wavelength-multiplexed_2014,
	year = {2014},
	month = {2},
	title = {Wavelength-multiplexed quantum networks with ultrafast frequency combs},
	volume = {8},
	rights = {2013 Springer Nature Limited},
	issn = {1749-4893},
	url = {https://www.nature.com/articles/nphoton.2013.340},
	doi = {10.1038/nphoton.2013.340},
	pages = {109--112},
	number = {2},
	journal = {Nature Photonics},
	shortjournal = {Nature Photon},
	publisher = {Nature Publishing Group},
	author = {Roslund, Jonathan and de Araújo, Renné Medeiros and Jiang, Shifeng and Fabre, Claude and Treps, Nicolas},
	urldate = {2025-01-14},
	date = {2014-02},
	langid = {english},
}

@article{cai_multimode_2017,
	year = {2017},
	month = {6},
	title = {Multimode entanglement in reconfigurable graph states using optical frequency combs},
	volume = {8},
	rights = {2017 The Author(s)},
	issn = {2041-1723},
	url = {https://www.nature.com/articles/ncomms15645},
	doi = {10.1038/ncomms15645},
	pages = {15645},
	number = {1},
	journal = {Nature Communications},
	shortjournal = {Nat Commun},
	publisher = {Nature Publishing Group},
	author = {Cai, Y. and Roslund, J. and Ferrini, G. and Arzani, F. and Xu, X. and Fabre, C. and Treps, N.},
	urldate = {2025-01-14},
	date = {2017-06-06},
	langid = {english},
}

@article{zhang_sub_2004,
	year = {2004},
	month = {2},
	title = {Sub femto-joule sensitive single-shot {OPA}-{XFROG} and its application in study of white-light supercontinuum generation},
	volume = {12},
	rights = {© 2004 Optical Society of America},
	issn = {1094-4087},
	url = {https://opg.optica.org/oe/abstract.cfm?uri=oe-12-4-574},
	doi = {10.1364/OPEX.12.000574},
	pages = {574--581},
	number = {4},
	journal = {Optics Express},
	shortjournal = {Opt. Express, {OE}},
	publisher = {Optica Publishing Group},
	author = {Zhang, Jing-Yuan and Lee, Chao-Kuei and Huang, J. Y. and Pan, Ci-Ling},
	urldate = {2024-11-22},
	date = {2004-02-23},
}

@article{serino_orchestrating_2024,
	year = {2024},
	month = {10},
	title = {Orchestrating time and color: a programmable source of high-dimensional entanglement},
	volume = {2},
	rights = {© 2024 Optica Publishing Group},
	issn = {2837-6714},
	url = {https://opg.optica.org/opticaq/abstract.cfm?uri=opticaq-2-5-339},
	doi = {10.1364/OPTICAQ.532334},
	shorttitle = {Orchestrating time and color},
	pages = {339--345},
	number = {5},
	journal = {Optica Quantum},
	shortjournal = {Optica Quantum, {OPTICAQ}},
	publisher = {Optica Publishing Group},
	author = {Serino, Laura and Ridder, Werner and Bhattacharjee, Abhinandan and Gil-Lopez, Jano and Brecht, Benjamin and Silberhorn, Christine},
	urldate = {2024-11-14},
	date = {2024-10-25},
}

@article{serino_realization_2023,
	year = {2023},
	month = {4},
	title = {Realization of a Multi-Output Quantum Pulse Gate for Decoding High-Dimensional Temporal Modes of Single-Photon States},
	volume = {4},
	issn = {2691-3399},
	url = {https://link.aps.org/doi/10.1103/PRXQuantum.4.020306},
	doi = {10.1103/PRXQuantum.4.020306},
	pages = {020306},
	number = {2},
	journal = {{PRX} Quantum},
	shortjournal = {{PRX} Quantum},
	author = {Serino, Laura and Gil-Lopez, Jano and Stefszky, Michael and Ricken, Raimund and Eigner, Christof and Brecht, Benjamin and Silberhorn, Christine},
	urldate = {2024-11-12},
	date = {2023-04-12},
	langid = {english},
}

@article{serino_orchestrating_2024-1,
	year = {2024},
	month = {10},
	title = {Orchestrating time and color: a programmable source of high-dimensional entanglement},
	volume = {2},
	rights = {© 2024 Optica Publishing Group},
	issn = {2837-6714},
	url = {https://opg.optica.org/opticaq/abstract.cfm?uri=opticaq-2-5-339},
	doi = {10.1364/OPTICAQ.532334},
	shorttitle = {Orchestrating time and color},
	pages = {339--345},
	number = {5},
	journal = {Optica Quantum},
	shortjournal = {Optica Quantum, {OPTICAQ}},
	publisher = {Optica Publishing Group},
	author = {Serino, Laura and Ridder, Werner and Bhattacharjee, Abhinandan and Gil-Lopez, Jano and Brecht, Benjamin and Silberhorn, Christine},
	urldate = {2024-10-30},
	date = {2024-10-25},
}

@article{shaked_lifting_2018,
	year = {2018},
	month = {2},
	title = {Lifting the bandwidth limit of optical homodyne measurement with broadband parametric amplification},
	volume = {9},
	rights = {2018 The Author(s)},
	issn = {2041-1723},
	url = {https://www.nature.com/articles/s41467-018-03083-5},
	doi = {10.1038/s41467-018-03083-5},
	pages = {609},
	number = {1},
	journal = {Nature Communications},
	shortjournal = {Nat Commun},
	publisher = {Nature Publishing Group},
	author = {Shaked, Yaakov and Michael, Yoad and Vered, Rafi Z. and Bello, Leon and Rosenbluh, Michael and Pe’er, Avi},
	urldate = {2024-08-21},
	date = {2018-02-09},
	langid = {english},
}

@article{obrien_optical_2007,
	year = {2007},
	month = {12},
	title = {Optical Quantum Computing},
	volume = {318},
	url = {https://www.science.org/doi/10.1126/science.1142892},
	doi = {10.1126/science.1142892},
	pages = {1567--1570},
	number = {5856},
	journal = {Science},
	publisher = {American Association for the Advancement of Science},
	author = {O'Brien, Jeremy L.},
	urldate = {2024-05-06},
	date = {2007-12-07},
}

@misc{yang_subcycle_2023,
	year = {2023},
	month = {7},
	title = {Subcycle tomography of quantum light},
	url = {http://arxiv.org/abs/2307.12812},
	number = {{arXiv}:2307.12812},
	publisher = {{arXiv}},
	author = {Yang, Geehyun and Kizmann, Matthias and Leitenstorfer, Alfred and Moskalenko, Andrey S.},
	urldate = {2024-04-04},
	date = {2023-07-24},
	eprinttype = {arxiv},
	eprint = {2307.12812 [physics, physics:quant-ph]},
}

@article{riek_subcycle_2017,
	year = {2017},
	month = {1},
	title = {Subcycle quantum electrodynamics},
	volume = {541},
	rights = {2017 Macmillan Publishers Limited, part of Springer Nature. All rights reserved.},
	issn = {1476-4687},
	url = {https://www.nature.com/articles/nature21024},
	doi = {10.1038/nature21024},
	pages = {376--379},
	number = {7637},
	journal = {Nature},
	publisher = {Nature Publishing Group},
	author = {Riek, C. and Sulzer, P. and Seeger, M. and Moskalenko, A. S. and Burkard, G. and Seletskiy, D. V. and Leitenstorfer, A.},
	urldate = {2024-03-21},
	date = {2017-01},
	langid = {english},
}

@article{wasilewski_pulsed_2006,
	year = {2006},
	month = {6},
	title = {Pulsed squeezed light: Simultaneous squeezing of multiple modes},
	volume = {73},
	issn = {1050-2947, 1094-1622},
	url = {https://link.aps.org/doi/10.1103/PhysRevA.73.063819},
	doi = {10.1103/PhysRevA.73.063819},
	shorttitle = {Pulsed squeezed light},
	pages = {063819},
	number = {6},
	journal = {Physical Review A},
	shortjournal = {Phys. Rev. A},
	author = {Wasilewski, Wojciech and Lvovsky, A. I. and Banaszek, Konrad and Radzewicz, Czesław},
	urldate = {2024-03-09},
	date = {2006-06-20},
	langid = {english},
}

@misc{ng_quantum_2023,
	year = {2023},
	month = {7},
	title = {Quantum noise dynamics in nonlinear pulse propagation},
	url = {http://arxiv.org/abs/2307.05464},
	number = {{arXiv}:2307.05464},
	publisher = {{arXiv}},
	author = {Ng, Edwin and Yanagimoto, Ryotatsu and Jankowski, Marc and Fejer, M. M. and Mabuchi, Hideo},
	urldate = {2024-03-05},
	date = {2023-07-11},
	eprinttype = {arxiv},
	eprint = {2307.05464 [physics, physics:quant-ph]},
}

@article{larsen_deterministic_2019,
	year = {2019},
	month = {10},
	title = {Deterministic generation of a two-dimensional cluster state},
	volume = {366},
	issn = {0036-8075, 1095-9203},
	url = {https://www.science.org/doi/10.1126/science.aay4354},
	doi = {10.1126/science.aay4354},
	pages = {369--372},
	number = {6463},
	journal = {Science},
	shortjournal = {Science},
	author = {Larsen, Mikkel V. and Guo, Xueshi and Breum, Casper R. and Neergaard-Nielsen, Jonas S. and Andersen, Ulrik L.},
	urldate = {2024-01-24},
	date = {2019-10-18},
	langid = {english},
}

@article{ralph_continuous_1999,
	year = {1999},
	month = {12},
	title = {Continuous variable quantum cryptography},
	volume = {61},
	issn = {1050-2947, 1094-1622},
	url = {https://link.aps.org/doi/10.1103/PhysRevA.61.010303},
	doi = {10.1103/PhysRevA.61.010303},
	pages = {010303},
	number = {1},
	journal = {Physical Review A},
	shortjournal = {Phys. Rev. A},
	author = {Ralph, T. C.},
	urldate = {2023-11-06},
	date = {1999-12-08},
	langid = {english},
}

@article{pirandola_advances_2020,
	year = {2020},
	month = {12},
	title = {Advances in quantum cryptography},
	volume = {12},
	rights = {© 2020 Optical Society of America},
	issn = {1943-8206},
	url = {https://opg.optica.org/aop/abstract.cfm?uri=aop-12-4-1012},
	doi = {10.1364/AOP.361502},
	pages = {1012--1236},
	number = {4},
	journal = {Advances in Optics and Photonics},
	shortjournal = {Adv. Opt. Photon., {AOP}},
	publisher = {Optica Publishing Group},
	author = {Pirandola, S. and Andersen, U. L. and Banchi, L. and Berta, M. and Bunandar, D. and Colbeck, R. and Englund, D. and Gehring, T. and Lupo, C. and Ottaviani, C. and Pereira, J. L. and Razavi, M. and Shaari, J. Shamsul and Tomamichel, M. and Usenko, V. C. and Vallone, G. and Villoresi, P. and Wallden, P.},
	urldate = {2023-11-06},
	date = {2020-12-31},
}

@article{kalash_wigner_2023,
	year = {2023},
	month = {9},
	title = {Wigner function tomography via optical parametric amplification},
	volume = {10},
	issn = {2334-2536},
	url = {https://opg.optica.org/abstract.cfm?URI=optica-10-9-1142},
	doi = {10.1364/OPTICA.488697},
	pages = {1142},
	number = {9},
	journal = {Optica},
	shortjournal = {Optica},
	author = {Kalash, Mahmoud and Chekhova, Maria V.},
	urldate = {2023-10-20},
	date = {2023-09-20},
	langid = {english},
}

@article{fabre_modes_2020,
	year = {2020},
	month = {9},
	title = {Modes and states in quantum optics},
	volume = {92},
	issn = {0034-6861, 1539-0756},
	url = {https://link.aps.org/doi/10.1103/RevModPhys.92.035005},
	doi = {10.1103/RevModPhys.92.035005},
	pages = {035005},
	number = {3},
	journal = {Reviews of Modern Physics},
	shortjournal = {Rev. Mod. Phys.},
	author = {Fabre, C. and Treps, N.},
	urldate = {2023-10-03},
	date = {2020-09-10},
	langid = {english},
}

@article{leverrier_unconditional_2009,
	year = {2009},
	month = {5},
	title = {Unconditional Security Proof of Long-Distance Continuous-Variable Quantum Key Distribution with Discrete Modulation},
	volume = {102},
	issn = {0031-9007, 1079-7114},
	url = {https://link.aps.org/doi/10.1103/PhysRevLett.102.180504},
	doi = {10.1103/PhysRevLett.102.180504},
	pages = {180504},
	number = {18},
	journal = {Physical Review Letters},
	shortjournal = {Phys. Rev. Lett.},
	author = {Leverrier, Anthony and Grangier, Philippe},
	urldate = {2023-09-12},
	date = {2009-05-06},
	langid = {english},
}

@article{zhong_quantum_2020,
	year = {2020},
	month = {12},
	title = {Quantum computational advantage using photons},
	volume = {370},
	url = {https://www.science.org/doi/10.1126/science.abe8770},
	doi = {10.1126/science.abe8770},
	pages = {1460--1463},
	number = {6523},
	journal = {Science},
	publisher = {American Association for the Advancement of Science},
	author = {Zhong, Han-Sen and Wang, Hui and Deng, Yu-Hao and Chen, Ming-Cheng and Peng, Li-Chao and Luo, Yi-Han and Qin, Jian and Wu, Dian and Ding, Xing and Hu, Yi and Hu, Peng and Yang, Xiao-Yan and Zhang, Wei-Jun and Li, Hao and Li, Yuxuan and Jiang, Xiao and Gan, Lin and Yang, Guangwen and You, Lixing and Wang, Zhen and Li, Li and Liu, Nai-Le and Lu, Chao-Yang and Pan, Jian-Wei},
	urldate = {2023-07-27},
	date = {2020-12-18},
}

@article{brecht_photon_2015,
	year = {2015},
	month = {10},
	title = {Photon Temporal Modes: A Complete Framework for Quantum Information Science},
	volume = {5},
	issn = {2160-3308},
	url = {https://link.aps.org/doi/10.1103/PhysRevX.5.041017},
	doi = {10.1103/PhysRevX.5.041017},
	shorttitle = {Photon Temporal Modes},
	pages = {041017},
	number = {4},
	journal = {Physical Review X},
	shortjournal = {Phys. Rev. X},
	author = {Brecht, B. and Reddy, Dileep V. and Silberhorn, C. and Raymer, M. G.},
	urldate = {2023-07-06},
	date = {2015-10-30},
	langid = {english},
}

@article{asavanant_generation_2019,
	year = {2019},
	month = {10},
	title = {Generation of time-domain-multiplexed two-dimensional cluster state},
	volume = {366},
	url = {https://www.science.org/doi/10.1126/science.aay2645},
	doi = {10.1126/science.aay2645},
	pages = {373--376},
	number = {6463},
	journal = {Science},
	publisher = {American Association for the Advancement of Science},
	author = {Asavanant, Warit and Shiozawa, Yu and Yokoyama, Shota and Charoensombutamon, Baramee and Emura, Hiroki and Alexander, Rafael N. and Takeda, Shuntaro and Yoshikawa, Jun-ichi and Menicucci, Nicolas C. and Yonezawa, Hidehiro and Furusawa, Akira},
	urldate = {2023-07-06},
	date = {2019-10-18},
}

@article{ledezma_intense_2022,
	year = {2022},
	month = {3},
	title = {Intense optical parametric amplification in dispersion-engineered nanophotonic lithium niobate waveguides},
	volume = {9},
	issn = {2334-2536},
	url = {https://opg.optica.org/abstract.cfm?URI=optica-9-3-303},
	doi = {10.1364/OPTICA.442332},
	pages = {303},
	number = {3},
	journal = {Optica},
	shortjournal = {Optica},
	author = {Ledezma, Luis and Sekine, Ryoto and Guo, Qiushi and Nehra, Rajveer and Jahani, Saman and Marandi, Alireza},
	urldate = {2023-06-27},
	date = {2022-03-20},
	langid = {english},
}

@article{nehra_few-cycle_2022,
	year = {2022},
	month = {9},
	title = {Few-cycle vacuum squeezing in nanophotonics},
	volume = {377},
	issn = {0036-8075, 1095-9203},
	url = {https://www.science.org/doi/10.1126/science.abo6213},
	doi = {10.1126/science.abo6213},
	pages = {1333--1337},
	number = {6612},
	journal = {Science},
	shortjournal = {Science},
	author = {Nehra, Rajveer and Sekine, Ryoto and Ledezma, Luis and Guo, Qiushi and Gray, Robert M. and Roy, Arkadev and Marandi, Alireza},
	urldate = {2023-06-27},
	date = {2022-09-16},
	langid = {english},
}

@article{cruz-rodriguez_quantum_2024,
	title = {Quantum phenomena in attosecond science},
	volume = {6},
	copyright = {2024 Springer Nature Limited},
	issn = {2522-5820},
	url = {https://www.nature.com/articles/s42254-024-00769-2},
	doi = {10.1038/s42254-024-00769-2},
	number = {11},
	urldate = {2026-03-04},
	journal = {Nature Reviews Physics},
	publisher = {Nature Publishing Group},
	author = {Cruz-Rodriguez, Lidice and Dey, Diptesh and Freibert, Antonia and Stammer, Philipp},
	month = nov,
	year = {2024},
	pages = {691--704}
}

@misc{hubenschmid_time-domain_2025,
	year = {2025},
	month = {6},
	title = {Time-domain field correlation measurements enable tomography of highly multimode quantum states of light},
	url = {http://arxiv.org/abs/2506.10483},
	doi = {10.48550/arXiv.2506.10483},
	number = {{arXiv}:2506.10483},
	publisher = {{arXiv}},
	author = {Hubenschmid, Emanuel and Burkard, Guido},
	urldate = {2025-06-16},
	date = {2025-06-12},
	eprinttype = {arxiv},
	eprint = {2506.10483 [quant-ph]},
}

@article{eto_quantum_2024,
	title = {Quantum frequency-resolved optical gating measurement for ultranarrow temporal correlation of twin beams},
	volume = {2},
	copyright = {© 2024 Optica Publishing Group},
	issn = {2837-6714},
	url = {https://opg.optica.org/opticaq/abstract.cfm?uri=opticaq-2-6-468},
	doi = {10.1364/OPTICAQ.533469},
	number = {6},
	urldate = {2026-03-05},
	journal = {Optica Quantum},
	publisher = {Optica Publishing Group},
	author = {Eto, Yujiro and Nuriya, Mutsuo and Kano, Hideaki},
	month = dec,
	year = {2024},
	pages = {468--474},
}

\end{document}